# A Systematic Literature Review on Task Allocation and Performance Management Techniques in Cloud Data Center

**Nidhika Chauhan[1], Navneet Kaur[2], Kamaljit Singh Saini[3], Sahil Verma[4], Abdulatif Alabdulatif[5], Ruba Abu Khurma[6], Maribel Garcia-Arenas[7] and Pedro A. Castillo[7,*]**

[1]University Institute of Computing, Chandigarh University Punjab, India, 143001.
[2,3]Department of Computer Science and Engineering, Chandigarh University Punjab, India, 143001.
[4]Department of Computer Science and Engineering, Uttaranchal University, Premnagar, Dehradun, Uttarakhand-248007, India.
[5]Department of Computer Science, College of Computer Qassim University, Buraydah, Saudi Arabia, 52571.
[6]MEU Research Unit, Faculty of Information Technology, Middle East University, Amman, 11831, Jordan.
[7]Department of Computer Engineering, Automatics and Robotics, University of Granada, Granada, Spain.
[*]Corresponding Author: Pedro A. Castillo. Email: pacv@ugr.es

**Abstract:** As cloud computing usage grows, cloud data centers play an increasingly important role. To maximize resource utilization, ensure service quality, and enhance system performance, it is crucial to allocate tasks and manage performance effectively. The purpose of this study is to provide an extensive analysis of task allocation and performance management techniques employed in cloud data centers. The aim is to systematically categorize and organize previous research by identifying the cloud computing methodologies, categories, and gaps. A literature review was conducted, which included the analysis of 463 task allocations and 480 performance management papers. The review revealed three task allocation research topics and seven performance management methods. Task allocation research areas are resource allocation, load-Balancing, and scheduling. Performance management includes monitoring and control, power and energy management, resource utilization optimization, quality of service management, fault management, virtual machine management, and network management. The study proposes new techniques to enhance cloud computing work allocation and performance management. Short-comings in each approach can guide future research. The research's findings on cloud data center task allocation and performance management can assist academics, practitioners, and cloud service providers in optimizing their systems for dependability, cost-effectiveness, and scalability. Innovative methodologies can steer future research to fill gaps in the literature.

## 1 Introduction

Cloud computing is an innovative and effective concept for offering Information and Communication technological services and tools on a pay-per-use basis. It is a utility framework in which on-demand network-dependent computing services are accessed through a pool of configurable computing resources [1]. Cloud Data center provides a cluster of resources like CPU, memory, RAM etc. through communicational links, cloud services and networks. The affordability of the cloud's pay-per-use approach has driven towards widespread adoption because it saves cloud service users the expense of acquiring and maintaining both equipment and software resources [2]. The key reason why numerous businesses are adopting cloud computing is that it is cost-effective, scalable, dependable, operates on a resource-sharing principle, is simple to use, and is secure. As the use of cloud applications grows, it becomes more difficult to control the whole request within the limited response time [3]. In addition to computational challenges, cloud computing faces issues such as server consolidation, workload distribution, Virtual Machine (VM) migration, energy utilization, and so on which majorly occur due to inefficient task allocation. The massive amount of data has made it challenging for data centers to efficiently distribute jobs to resources while maintaining Quality of Service (QoS) and a cloud service provider's profitability [4]. As a result of the aforementioned problems, an efficient scheduling method is required [5-7]. Scheduling is used to manage user requests by allocating appropriate resources and balancing the load across VMs. It is performed by virtualization, which divides a single physical system or server into numerous virtual machines (VM). Multiple tasks are assigned to each virtual machine [8]. Because of the expanding usage of cloud servers, virtualization is becoming increasingly important in managing numerous servers on a similar shared infrastructure. While virtualization technology offers cost savings, it also increases system complexity. As a result, virtual machines perform poorer than actual host machines doing the same workload. To address this issue, Cloud Service Providers (CSPs) are given the role of allocating incoming tasks to the appropriate VM, ensuring that no machine is overburdened and that the load is distributed evenly across various machines or resources. It is also used to increase QoS parameters such as task resignation proportion, resource consumption, reliability, resource utilization, computational expenses, least time complexity with limited makespan, and throughput this eventually will enhance the performance of the cloud data center [9]. With increasing demand in QoS, the performance of the system also gets affected and so is the users' interest towards the cloud services. In general, performance management is concerned with the actual performance of hardware or a virtual system. It considers factors such as workload, CPU and memory utilization and system latency.

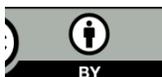

The evaluation of various metrics and benchmarks for cloud systems falls under the purview of cloud performance management. It is applied to examine how well a cloud system is operating and the scope for enhancements. To ensure the credibility of any application, its performance needs to be evaluated and compared with existing applications or techniques.

A systematic literature review is required to offer a comprehensive and organised summary of the research in the field, considering the numerous proposed methodologies. A literature review is conducted to comprehensively evaluate and synthesise existing knowledge on cloud data centre work allocation and performance management strategies. The review aims to identify trends, research gaps, strengths, and weaknesses in the field. Undertaking a systematic literature study enables researchers and practitioners to acquire knowledge on state-of-the-art approaches, methodologies, and performance evaluation measures [8]. A critical evaluation can also unveil domains for additional investigation, accentuate developing patterns, and aid in the comparison and selection of methodologies based on application prerequisites. A comprehensive analysis of scholarly literature on task allocation and performance management techniques in cloud data centres facilitates comprehension of the present research scenario, provides direction for future research, and assists professionals in selecting optimal methods for enhancing task allocation and performance management.We present a detailed systematic literature review to better understand the research trends and challenges in task allocation and performance management strategies in cloud data centers. The work focuses primarily on how research articles that employ task allocation and performance management methodologies leverage their capabilities in managing user requests, allocating workloads and contributing to performance management. Furthermore, the purpose of this paper is to investigate research endeavors related to task allocation, the role of performance, and its evaluation in cloud data centers. The study was conducted methodically, and articles were selected regarding (i) task allocation issues, (ii) the various techniques for task allocation in a cloud environment, (iii) the role of performance and its evaluation in cloud data center and (iv) the parameters that remained unaddressed concerning performance and task allocation in data centers.

The rest of the paper is organized as follows: There is a brief literature review in Section 2. Section 3

contains the study design for a systematic literature review survey, including research objectives, information sources, and inclusive and excluding requirements for research articles. Section 4 examines cloud task allocation and different job allocation approaches. Section 5 explores several ways to cloud data center performance management. Section 6 goes over the findings and discussions. Section 7 addresses outstanding concerns.

## 2. Literature Review

Many researchers have investigated and demonstrated the impact of various task allocation and performance techniques in cloud data centers, but few have conducted systematic literature reviews in these areas. Furthermore, to the best of the author's knowledge, no study has combined task allocation and performance management techniques. Several landmarks for task allocation and performance management have been provided by authors such as Banga et al. [10], Rodriguez and Buyya [11], Weerasiri et al. [12], Khallouli and Huang [13], Singh and Chana [14,15], Jennings et al. [16], Jiang [17], Mann et al. [18], and Navimipour [19].

Banga et al. [10] recommended a conventional cost-based scheduling approach that assigns suitably preferred resources, lowering the total cost of implementation and operation. Several tasks/cloudlets have also been divided and assigned appropriate resources to complete the task in the shortest possible time and at the lowest possible cost, based on their computational capability. Rodriguez and Buyya [11] conducted research on modern cloud computing technologies. They recommended classifying these systems according to different workload kinds, architectures, levels of complexity, and objectives. This research examined the 10 organizational business frameworks "Borg, Kubernetes, Swarm, Mesos, Marathon, Yarn, Omega, Apollo, and Fuxi". Additionally, Weerasiri et al. [12] looked at resource management tools (i.e., cluster control systems in charge of container-based task development and monitoring) applied in massive production clusters. Based on their operating system (e.g., "public, private, hybrid, virtualization-based, container-based"), cloud environment (IaaS, PaaS, or SaaS), and scheduling approach, they created a detailed taxonomy of resource strategies (e.g., regulation or user-specific). The study's main emphasis was on several deployable open-source techniques. The investigation provided comprehensive perspectives on container-based workflow techniques that only touched on a few processes. Furthermore, the survey made no mention of the current cluster scheduling framework. Multi-perspective research that covers several facets of the job allocation problem is recommended by Khallouli and Huang [13]. As the complexity of the original problem increases, resource planning is a prominent topic in cloud computing, and research in this field is continually advancing. Two significant studies on resource provisioning for cloud computing and work schedules were carried out by Singh and Chana [14-15]. Concerns with dynamic resource management in cloud computing were examined by Jennings et al. [16]. Numerous more studies focused on certain schedules or techniques for using resources (such as workflow (DAG) scheduling algorithms). Jiang [17] carried out a study in which he assessed and evaluated works on task allocation and load-Balancing. The analysis is focused on the general features of diverse distributed systems and investigates the approaches in terms of several elements, including control models, resource optimization methods, methods for attaining dependability, coordination mechanisms among heterogeneous nodes, and models that account for network topologies. Based on these characteristics, a full taxonomy is developed and described, although this study is not focused on VM scheduling and cloud architecture. Mann et al. [18] investigated the issue of inefficient use of physical resources and the VM allocation challenge in cloud data centers. The research looked at frameworks and algorithms for a variety of scenarios. They studied the most commonly used problem formulations and recommended a more holistic assessment of the current state of research in VM allotment, but the study does not focus on VM load-Balancing. Load-Balancing is a prominent concern in cloud data centers, according to Milani Navimipour [19]. It is important to prevent over- and under-utilizing resources, which may be done by using load-balancing techniques that are successful. The selection of an acceptable resource for a certain work does not guarantee that the resource will stay optimized during the job's execution; it may overburden the PM at some point. To fully explore this problem, the authors carried out an organized literature study on cloud load-balancing techniques.

Numerous algorithms with benefits, downsides, and obstacles have been examined, and the study has been carefully categorized. According to the author's analysis of reaction times and costs, hybrid algorithms are advantageous for both cloud users and cloud developers.

Ramezani et al. [20] developed an optimization framework that used Particle Swarm Optimization (PSO) as a task scheduling model to migrate the additional jobs to the new host VMs. The outcomes demonstrated that the load-balancing method was quicker than conventional load-balancing algorithms. Reduced VM downtime and the possibility of users losing vital data were the main goals of the study. This led to lower memory use, lower migration and makespan, and increased data center efficiency, which enhanced the QoS that cloud users experienced. Although the approach emphasizes autonomous activities and homogeneous virtual machines, it has limited scalability.

## 2.1 Task Allocation

The development of green cloud computing can be facilitated by addressing task allocation in the cloud computing environment. Task allocation is a significant research problem due to its impact on resource response times, energy consumption, and system costs. Suboptimal task allocation can result in elevated makespan and increased overall costs, emphasizing the need for efficient allocation algorithms. Flexibility, dependability, resource utilization optimization, and achieving maximum throughput are crucial considerations in task allocation. The allocation of tasks directly affects CPU and memory utilization and Quality of Service (QoS) metrics, highlighting the importance of proficient allocation algorithms. In modern computer systems, task allocation is influenced by task arrival patterns and associated energy consumption, aiming to minimize energy usage and enhance overall energy efficiency. Resource management and planning play a critical role in task allocation, as rapid decision-making for low-latency operations is vital for performance optimization.

Effective resource planning solutions are essential to ensure efficient task allocation and resource availability. Job and resource scheduling are key aspects of task allocation, where cost-effectiveness, flexibility, and meeting deadlines are crucial considerations. Effective resource allocation strategies are necessary to address the challenges of managing dynamic resources and optimize performance factors such as cost, scalability, and manageability. Load-Balancing is a crucial methodology in task allocation, as resource imbalance can negatively impact system performance. Efficient load-Balancing minimizes task execution and data transmission time, optimizes RAM utilization, and reduces migration time and makespan, thereby enhancing data center efficiency. Task allocation for unmanned aerial vehicles (UAVs) presents unique challenges, considering task arrival patterns and energy utilization in UAV-based environments. Innovative strategies are required for efficient task allocation in this emerging area. Dynamic workload migration is a critical factor in task allocation, considering power consumption and cost. Efficient workload distribution and adaptation to dynamic data center demands require proper consideration of these factors. Task scheduling significantly influences resource utilization, and optimizing CPU and memory utilization through proper process scheduling enhances overall resource efficiency in the data center. Inadequate resource allocation in cloud environments can lead to increased power consumption and resource utilization duration, impacting data center performance.

Developing efficient and effective resource allocation methodologies is essential. By addressing these issues, enterprises can maximize resource utilization, boost performance, and achieve efficient task allocation in their data center environments. Tab. 1 depitcs issues related to task allocation and its relation with the performance of the data center.Zhou et al. [21], Rahman et al. [22], Krishna [23], Malini and Navimipour [19], and Akintoye and Bagula [24] investigated various task allocation challenges and identified features such as task execution costs and response time. According to Zhou et al. [21], as the use of cloud applications grows, there is a greater need for effective task allocation on computing machines. With large data centers, efficiently allocating tasks to resources while keeping QoS and profitability of a cloud service provider in mind has become a challenge. According to Rahman et al [22], Krishna [23], Malini and Navimipour [19], to overcome the task allocation challenge, multiple approaches are being investigated that can improve task allocation and ensure that all operations are completed with greater efficiency. These approaches consider various objectives for allocation decisions, such as quick response time, lower energy consumption, overall user satisfaction, and cloud environment performance. According to Akintoye and Bagula [24], task allocation is a critical feature that must be incorporated into all cloud-based computing platforms so that tasks are properly allocated and resources are managed efficiently to provide the best possible experience to cloud users. It has a significant impact on the allocation of resources for new activities with performance optimization constraints. Outstanding quality, profitability, consumption, adaptability, delivery convenience, and affordability are all key goals for any cloud computing environment that can only be achieved through effective task allocation.

Table 1: Issues related to task allocation and its relation with the performance of the data center.

| S. no | Author | Issues | Performance Aspects | Solutions |
|---|---|---|---|---|
| 1 | Banga et al. [10] | Inefficient task allocation | Makespan and Increased cost | Cost-based scheduling approach, dividing tasks and assigning appropriate resources. |
| 2 | Rodriguez and Buyya[11] | Flexibility and reliability | CPU, Memory Utilization, and QoS | Research on modern cloud technologies, business frameworks. |
| 3 | Weerasiri et al. [12] | Utility-based computing | Task arrival and energy consumption | Resource management tools, taxonomy of strategies. |
| 4 | Khallouli and Huang [13] | Resource planning | Fast scheduling decisions | Emphasized multi-perspective research. |
| 5 | Singh and Chana [14] | Resource provisioning | QoS, execution time, energy utilization, and scalability | Research on resource provisioning and work schedules. |
| 6 | Singh and Chana [15] | Job and resource Scheduling | Cheap and flexible, deadline constraint | Genetic algorithm-based approach for scheduling. |
| 7 | Jennings et al. [16] | Dynamic resource management | Cost, scalability, and manageability | Decentralized approach for resource management. |
| 8 | Jiang [17] | Task Allocation and load balancing | Improved coordination abilities | Extensive analysis of task allocation and load balancing. |
| 9 | Mann et al. [18] | VM allocation in cloud data centers | Monetary costs and environmental impact | Investigation of VM allocation challenges. |
| 10 | Milani and Navimipour [19] | Load-Balancing | Resource over/underload | Systematic literature study on cloud load-balancing techniques. |
| 11 | Ramezani et al. [20] | Load-Balancing | Low task execution and transferring time, less memory utilization, low migration, and Makespan | Optimization framework using PSO for load-balancing. |

| 12 | Zhou et al. [21] | Task allocation and offloading to UAV assigned clouds | Task arrival and energy consumption | Task allocation and offloading to UAVs in cloud environments. |
|---|---|---|---|---|
| 13 | Rahman et al. [22] | Dynamic workload migration | Power consumption and cost | Addressed dynamic workload migration in cloud environments. |
| 14 | Krishna [23] | Task scheduling affects resource utilization | CPU and Memory Utilization | Explored how task scheduling influences resource utilization. |
| 15 | Akintoye and Bagula[24] | Difficulties in resource allocation | Performance impact due to resource use and energy consumption | Addressed difficulties in resource allocation, considering performance impact. |

## 2.2. Algorithms for Type of Task Allocation

**Heuristic Algorithm**

Gawali and Shinde [25] investigated heuristic approaches, which are a set of rules derived from prior experiences that aim to solve problems more quickly than traditional methods. The main goal is to find the estimated solution to a problem as quickly as possible, and the technique employs a variety of shortcuts and methods to generate an optimal solution in a short amount of time. According to Hussain and Bagh [26], heuristic algorithms have the advantage of providing an efficient solution within the cost and timeframe constraints. They are simple to set up and run quickly, making them an ideal environment for online task scheduling. Mor et al. [27] stated heuristic algorithms work on NP-hard task scheduling problems, multi-resource allocation, and heterogeneous environments.

**Meta Heuristic Algorithm**

According to Jain and Upadhyay [28], metaheuristic algorithms are inspired by nature and developed to solve general problems. They adhere to a set of guidelines to fix the issue. The algorithms take a long time to run since there is a significant amount of solution space to consider when evaluating the final result. According to Singh and Kumar [29], metaheuristic algorithms are stochastic processes that take time to merge, and the outcomes of the solutions depend on the problem's nature, its early setup, and the approach used to look for answers. Rodrigues et al. [30] stated that the traditional and conventional optimization techniques are still struggling to efficiently deal with NP-hard problems, non-linear, high dimension, and hybrid issues. Metaheuristic algorithms seem to be the preferred solution for these problems, and they are widely used in many domains.

**Hybrid Algorithm**

Dubey and Sharma [31] stated that hybrid algorithms are a combination of heuristic and metaheuristic approaches. Heuristic methods are used for initial VM placement, and metaheuristic techniques are used to optimize VM placement during migration. According to Mohanty et al. [32], a metaheuristic technique may be used to develop a collection of solutions, and a heuristic approach is then used to choose the best option. When time, money, and available options are taken into account, both solutions appear advantageous, but as soon as execution gets underway, complexity rises.

## 2.3. Performance Management

Performance management, according to Mehrotra and Srivastava [33], is the practice of analyzing multiple matrices and benchmarks in a cloud system. Any application's performance has to be assessed and contrasted with those of other apps or methods to ensure its credibility. According to Chen and Li [34], the dynamic nature of the cloud model makes resource management in the cloud difficult. The performance metrics analyze the attributes to understand the factors that affect the performance of the cloud environment.

**Performance Metric**

The response time, the overall number of SLA violations, and the amount of power used by the resources to fulfil user requests, according to Zareian et al. [35], are used to assess the performance management of cloud frameworks. Gupta et al. [36] state that it is a common practice to employ particular performance metrics and characteristics as the goal function to reduce the detrimental impact on QoS. Song [37] suggests that the information provided by performance measurements may be useful for locating the true origin of a breakdown that propagates across the system's various levels.

**Performance Management Approaches**

According to Ganesh et al. [38], performance management strategies are distinct and have different metrics when it comes to cloud computing. Performance management, according to Fareghzadeh et al. [39], focuses on anomalies in dynamic systems. These performance issues are regarded as exceptions and variances in the system, which are recognized using a variety of techniques, such as Dynamic Allocation, Power-Aware Approach, Model-Based Approach, Rule Base Approach, Unsupervised and Supervised Approach, Resource Capacity-Based Approach, and Scaling Approach. According to Van et al. [40], the performance management approach also offers application latency optimization in cloud forms, which is effective in determining the scalability limit of the activities.

## 3. Research Method

The process of doing a literature review includes creating research questions, describing the collection of datasets to be examined, collecting data, evaluating data, and summarizing findings. The systematic literature review's main goal is to draw attention to gaps that previous research has found.

### 3.1. Review Methodology

The review approach makes use of the following steps: creating research questions, selecting the datasets to be examined, and reviewing observations and conversations. The process comprises examining main and secondary databases, applying eligibility criteria for inclusion and exclusion, implementing dynamic requirements, and summarizing with discussions. Every author's study is carefully examined to keep the procedure development realistic. When there is conflict, the viewpoints are freely discussed and decided through an iterative process and observations. Both electronic and manual datasets, including the most esteemed journals, conference proceedings, and researcher dissertations, are examined for the review of the literature. Overall, 480 articles were identified during the initial search; however, only 118 were ultimately picked when the approach was used.

### 3.2. Research Question

This systematic literature review's goal is to present the most recent findings on work distribution and its methods, as well as on performance management strategies and measurements. To undertake a systematic review, the following set of research questions given in Tab.2 has been established.

Table 2. Research Questions.

| Research Question | Description |
|---|---|
| RQ.1 | What are the key issues associated with task allocation and how do they impact the performance of data centers? |

| Research Question | Description |
|---|---|
| RQ.2 | What are the diverse techniques utilized for task allocation in cloud environments? |
| RQ.3 | What is the role and significance of performance evaluation in the context of cloud data centers? |
| RQ.4 | Which parameters pertaining to performance and task allocation in data centers remain unaddressed? |

### 3.3 Sources of Information

To conduct a systematic review, a larger perspective is required. So before conducting the review, an appropriate database should be selected that could effortlessly offer suitable outcomes as per the pertinent keywords.

The following five databases were considered:

- Springer (http://www.springer.com/in/)
- Scopus (http://www.scopus.com)
- IEEE eXplore (http://ieeexplore.ieee.org/)
- Web of Science (https://www.clarivate.com)
- ScienceDirect (http://www.sciencedirect.com/)

Additional Sources of Information

- Book Chapters
- Papers in Conferences
- Technical Reports
- Thesis

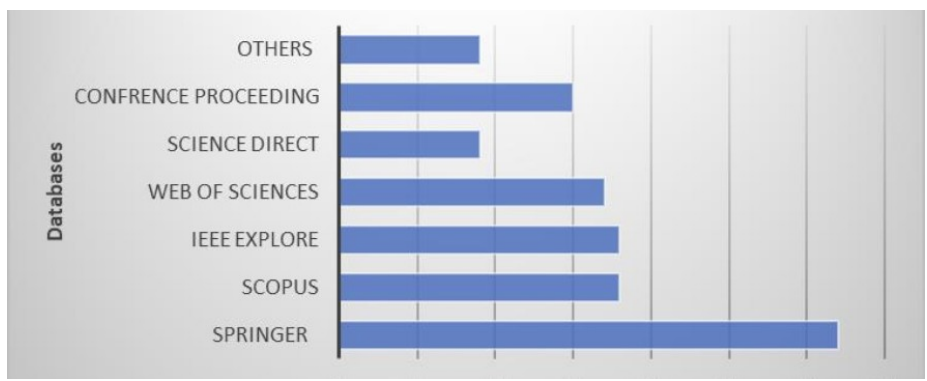

**Figure 1.** Search Database.

Fig. 1 provides a visual representation of the databases that were included in the search process. The graph clearly shows the specific databases that were searched to gather relevant information for the study or analysis. By presenting this information graphically, it allows for a quick and easy understanding of the databases that were considered in the research.

### 3.4 Important Search Keywords

All the primary and additional datasets for the survey were searched for the given set of keywords. The search was conducted between 2012 to 2022. The keywords used are mentioned below in Fig.2.
- Task Allocation
- Cloud Data Centres
- Performance Management
- Task Allocation Approaches
- Performance Matrices
- Performance Management Approaches

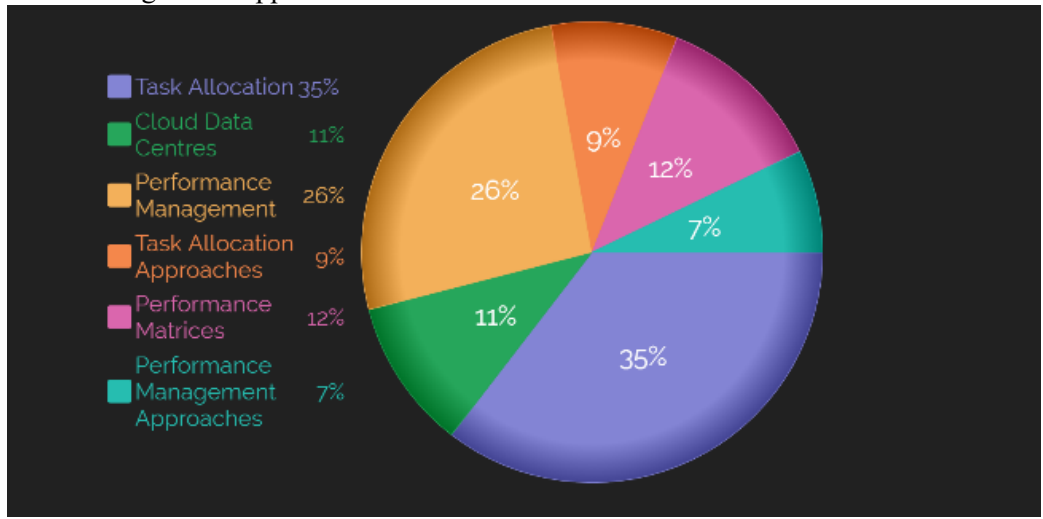

**Figure 2.** Article Search.

### 3.5 Criteria for Inclusion and Exclusion

Three phases of adaptation were made to the inclusion and exclusion criteria, with completely unrelated publications being eliminated from the search. Only publications dealing with computer science and engineering were included. However, given the multidisciplinary nature of terms like "task allocation" and "performance," publications from numerous contexts were found when searching. All of the additional publications cover a variety of topics including blockchain, microservices, management, networking, and healthcare. These papers were dismissed. Only papers written in English were taken into consideration. The systematic review comprised research publications published between 2012 and the present. The identical research articles from various sources have also been eliminated to increase the search's credibility, while related research papers written by the same authors with minor changes are still taken into account. Consideration is also given to works that have been presented at conferences and subsequently published in respectable journals. The systematic literature review is depicted in Fig. 3 below. The review is divided into the following three stages: In the initial phase, we found 943 documents. In the second round, the search was reduced to 480 papers based on their titles. Additionally, 267 articles in the third round were considered based on the abstract. Finally, 118 publications are selected using inclusion and exclusion criteria from the complete dataset.

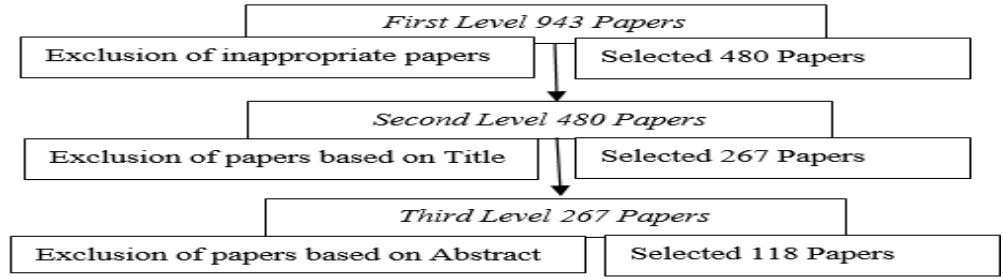

**Figure 3.** Exclusion Criteria for Search

### 3.6 Quality Assessment

The next procedure is to do a quality assessment of the chosen papers after the inclusion and exclusion criteria are completed. There were several articles and conference papers linked to our area of interest because computer science is such a vast field. The assessment of the quality of selected papers was done as per CRD guidelines which stated that each study was to be assessed for bias, and internal and external validation outcomes [41]. Based on the criteria listed in Appendix A, the evaluation is conducted. Once a paper was chosen, it was evaluated for the classification criteria mentioned in Part 2 and, if the result was positive, part 3 and Part 4 of the assessment were completed. Part 1 of the appendix is used to evaluate the rationale for screening.

### 3.7 Data Extraction

In the systematic review, data were taken from 118 research publications. The papers were chosen using the inclusion-exclusion criteria from the top sources listed in Tab. 3. These articles were chosen from 2012's digital libraries. The following steps are taken in the data extraction process:

• The 118 articles were all reviewed, and the issue under investigation was focused on.

• Review articles from 2012 to the present were taken into consideration.

• The works that appeared in prominent journals were taken into account.

**Table 3.** Databases searched

| S.no | E-Resource | Keywords | Year of search | No of papers | Paper Type |
|---|---|---|---|---|---|
| 1 | Springer | Task Allocation, Cloud data centre | 2012-present | 32 | Review, Implementation, Case studies |
| 2 | Scopus | Cloud data centre | | 18 | |
| 3 | IEEE eXplore | Performance Management | | 18 | |
| 4 | Web of Sciences | Performance Metrics Performance | | 17 | Survey |

**RQ1: What are the issues related to task allocation and how is it related to the performance of the data center?**

### 4. Task Allocation in Cloud Environment

As the usage of cloud applications grows, effective task allocation on computing machines becomes extremely important. It has become difficult to distribute tasks to resources effectively while considering the Quality of Service and a cloud service provider's profitability in light of big data centers [42]. These methods take into account a variety of factors while making allocation decisions, including rapid response times, low energy usage, and cloud environment performance [23,43,44]. Based on these objectives, task allocation algorithms under study are categorized into Heuristic, Meta-Heuristic, and Hybrid algorithms as shown in Fig.1. The selection process adopted for categorization of algorithms into Heuristic, Meta-Heuristic, and Hybrid algorithms as shown in Fig. 1 is done through considering keywords and abstracts. Most of the algorithms were rejected because they took into account the energy consumption of servers, CPUs, or memories without considering the energy consumption of networking resources which is a significant parameter of this study. The few studies that looked at network resource algorithms did not consider carbon emissions, task allocation, and energy efficiency as key factors in data centers. For the relevant algorithm selection, the inclusion and exclusion techniques are applied.

**RQ 2: What are the various techniques for task allocation in the cloud environment?**

**Task Allocation Techniques**

The allocation of tasks in cloud data centers is optimized through the use of heuristic, meta-heuristic, and hybrid methodologies. Heuristic methods employ pre-established standards to allocate resources through either rule-based or intuitive methodologies. Metaheuristic algorithms employ iterative search methods that are inspired by natural phenomena to identify nearly optimal solutions. This is in contrast to other types of algorithms. Hybrid methodologies combine heuristic and metaheuristic techniques to achieve a balance between computational efficiency and accuracy in task allocation. The implementation of these techniques shown in Fig. 4 enables efficient utilization of resources, improved performance, and enhanced scalability in cloud environments by dynamically assigning tasks to suitable resources based on workload, resource availability, and task requirements.

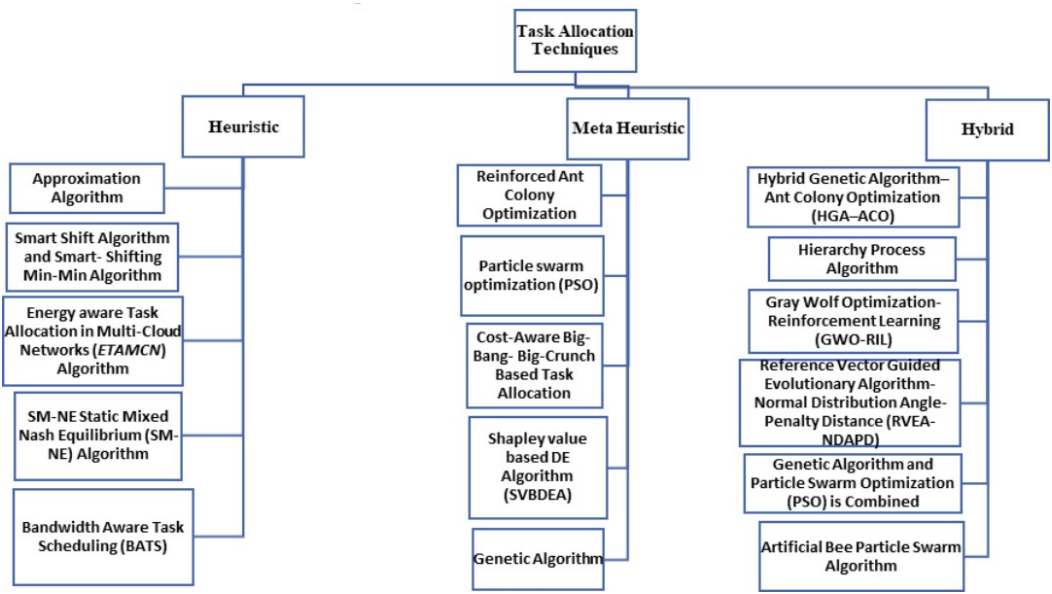

**Figure 4.** Classification of Task Allocation Techniques.

**Heuristic Algorithm**

Heuristic methods attempt to solve problems faster than traditional methods, with the primary goal of finding an approximated solution. They are a set of laws derived from previous encounters. Using a variety of shortcuts and procedures, the approach generates an effective solution within budget and in a limited amount of time [45]. They are regarded as an appropriate setting for online task scheduling where they are expected to provide a quick resolution because they are simple to use and operate more quickly [46]. The heuristic methods that satisfy the constraints of QoS, SLA, cost, amount of migrations, and performance time were studied and are depicted in Fig. 4. In an online setting, they are anticipated to offer a swift resolution [46]. The reason for selecting these algorithms is because they provide the solution for NP-hard task scheduling problems, multi-resource allocation and heterogeneous environments.

**Approximation Algorithm**

This approach is used to manage an optimization problem's NP-completeness. Decision-related issues are covered by NP-completeness. The NP set's most difficult puzzles are these. The strategy can deal with complicated issues, but it cannot guarantee that the solution found will be the optimal one. The basic goal of an approximation algorithm is to deliver a solution that is as close as possible to the anticipated ideal answer while still being computationally efficient. A few of the research, which will be further addressed, employed an approximation strategy to assign tasks [47]. To consume less total energy, Liu et al. [48] analyze the job allocation difficulties in a diverse resource environment. The effort is specifically taken into account for configurations that effectively offload binary computing after a job is completed. Task allocation and offloading choice are two subtasks of the integer programming problem of binary computation offloading. These subproblems are resolved using the approximation algorithm. The majority of smart devices employ a shared network channel with unexpected transmission medium changes and MEC (Mobile Edge Computing) servers, which have constrained resources and don't fully satisfy all criteria. If the MEC server is offloading multiple mobile device activities, the uploading procedures may cause wireless disruption. Energy use will rise and data transmission times will be extended. A restricted frequency subchannel is taken into consideration for this issue. To increase the efficiency of wireless access, only a small number of mobile devices can send data simultaneously in this subchannel. Task completion time is used as a parameter along with the sub-channel. A strong NP-hard issue is the job distribution problem without taking deadlines into account.

**Smart Shift Algorithm and Smart-Shifting Min-Min Algorithm**

The Smart Shift (SmS) Algorithm is used to perform shifting operations to reduce task allocation time or energy costs. The objective of this algorithm is to determine a reason for the Greedy approach to create a non-optimal solution. In scenarios where the Greedy approach allocates a non-optimal solution, the assigned processing unit may be inefficient in comparison to other unselected processing units because they are already in use for other processes. The Greedy algorithm assigns unselected processing units based on task priority. The objective of SmS approach is to avoid the non-optimal solution problem [49]. In a cloud environment, the "Smart Shifting Min Min (SmS-MM) Algorithm" is used to create a task allocation plan. This algorithm has two main components i.e., Smart Shift and Min-Min manipulations. The previous paragraph discussed the Smart Shift. "Min-Min is an optimization technique" that considers two methods (SmS-MM) for minimizing a pattern. These approaches are used in this study to reduce execution time and energy costs [50]. Gai et al. [49] investigated the task allocation problem in cloud, edge, and fog computing and proposed a novel approach for resource management called "Energy-aware

Fog Resource Optimization" (EFRO). The data transfer over the cloud is one of the reasons that encourages traffic jams thus affecting the QoS or connecting requests. This becomes a major issue for enterprises which require large simulation services from time to time. The heavy workload increases energy consumption and task execution time, affecting work efficiency and energy savings. The algorithm is used to reduce energy costs as well as time consumption. It uses standardization and small shift operations to find the optimal resource allocation solution. The model focuses on two main factors i.e., task assignment and energy saving. The considered environment is heterogeneous, and all servers have been considered. The task forwarding is handled dynamically based on workload. There are three components in the model i.e., cloud, fog, and edge. The model takes into account each device while keeping their computational unit in mind. The aim is to reduce total energy in the allotted time. For the model implementation, the "Standardization Table Creation (STC) algorithm" is used for creating an intermediate table for task assignment, the SmS algorithm is used for sub-optimal solution and the SmS-MM algorithm is used for holistic control for the task allocation plan. Under a variety of workload circumstances, EFRO attains improved energy and plan-generation efficiency by reducing the time spent on generating allocation plans.

**Energy-aware Task Allocation in Multi-Cloud Networks (ETAMCN) Algorithm**

Task allocation was identified as a challenging problem in a multi-heterogeneous cloud by Mishra et al [51]. The authors believed that task allocation was the best option for improving performance parameters such as energy utilization and makespan. As a result, the authors developed a task-based allocation algorithm called ETAMCN. The algorithm shortens the makespan and reduces energy consumption. The work is divided into two phases. In Phase 1, the VM selection is done using an Expected Time to Complete (ETC) matrix. In Phase 2, VMs with lower energy consumption are chosen using an Energy Consumption (EC) matrix, assuming that the VM has sufficient resources for task execution. The network receives verified tasks, which are then queued with the help of the Max Heap data structure. The tasks are stored in the order of their priority, which is determined by task length divided by task deadline. The cloud manager's role is to control the task queue and monitor the VMs on the network. To assign a single task, a set of VMs is searched considering that the VM can perform the execution by maintaining SLA. The ETC matrix is used to search for VMs, and the EC matrix is used to select the VMs that use the least amount of energy. For the input values, EC and execution time are taken into account, and the input values are updated each time a request is shared. When the task is finished, CSP releases the resource and updates the system.

**SM-NE Static Mixed Nash Equilibrium (SM-NE) Algorithm**

According to Josilo and Dan [52], resource management is a challenging issue in fog computing because the number of offloading options increases as the number of devices increases. According to the authors, developing a low-complexity algorithm is a difficult task. As a result, they created a model based on game theory and inequality theory for allocating computational tasks to nearby devices. An algorithm known as AM-NE for resource allocation for task execution on nearby devices and edge cloud was proposed based on equilibrium techniques. In this system, devices can choose whether to offload the computation to a device closer to them or the edge cloud; however, the model relied solely on average system parameters, whereas other performance parameters such as energy efficiency, markspan, and so on should also be considered.

**Bandwidth Aware Task Scheduling (BATS) and Bar Optimization Algorithm**

BATS algorithm was introduced by Lin et.al [53] for task scheduling and resource management. The algorithm developed a nonlinear programming approach to address issues with dividing tasks. The system focuses on independent tasks of the same size but when resources are allocated the system does not consider VM load which results in long waiting time and execution time. Acedo and Rosa [54] pioneered bar optimization. The method investigates the social behaviour of bartenders when dealing with customer orders. Customer request handling is regarded as an NP-hard scheduling problem, with the main parameters being serving time, serving cost, order selection criteria, and so on. The environment in which a bartender works, and in which he is expected to make decisions, is extremely dynamic, time-sensitive, and asynchronous. When compared to other Greedy strategies, the optimization produced a good performance outcome. Gawali and Shinde [25] discovered that to improve cloud performance, task allocation and resource scheduling issues must be addressed. To accomplish this, the author introduced a technique that employs the "Modified Analytic Hierarchy Process (MAHP), BATS + Bar optimization, Longest Expected Processing Time pre-emption (LEPT)," and the divide and conquer technique. Every task in this work is processed before being assigned to a resource. For resource allocation, BATS + Bar optimization is used. LEPT is used for task pre-emption. The results indicate that the resource utilization has improved in this system but the parameters like turnaround time and response time have been ignored.

**Meta Heuristic Algorithm**

These algorithms were created to address common issues and were inspired by nature. They adhere to a set of guidelines to fix the issue. The algorithms take a long time to run since there is a significant amount of solution space to consider when evaluating the final result. These algorithms are stochastic processes that take time to merge, and the solutions depend on the type of issue at hand, how it was set up in advance, and how it was discovered. Conventional optimization strategies continue to struggle to effectively solve NP-hard difficulties, non-linear, large dimension, and hybrid problems. Metaheuristic algorithms are often employed in many different fields and appear to be the preferable answer for these problems [28]. Several meta-heuristic algorithms are discussed further below. These algorithms are chosen for their task scheduling, load-Balancing, security, scalability, virtualization, and recovery capabilities.

**Reinforced Ant Colony Optimization based Task Allocation**

The Ant Colony Optimization (ACO) algorithm was inspired by ants. According to ant behaviour, they begin their food search by moving randomly, and once the search is completed, they return to their colony, leaving a pheromonal trail. Other ants are likely to follow that pheromonal trail if they find it. The pheromone trail fades over time as the intensity decreases. After some time, the ants search for a shorter path, and the path becomes more frequent as more ants travel through it, and thus the pheromone density increases on this path. Because ants emit pheromones while travelling, the shortest path becomes the best solution [55]. Reinforcement learning (RL) is a branch of Machine Learning. It is concerned with how software agents must act to enhance reward based on the specific situation, i.e., attributing a benefit to the current solution to achieve the maximum solution. RL employs several software programs to determine the best possible actions or paths to take in each situation [56]. Nalini and Khilar [57] identified that task allocation in VM is a difficult job and increased bandwidth requirement by cloud services is a major issue. It is a combinatorial optimization issue which needs to be focused upon. To overcome this problem task scheduling is necessary. The author considered makespan as a primary objective because it gets affected by other factors like the number of tasks, execution time, task scheduling and fault tolerance. According to Nalini and Khilar [57], task allocation in VM is a difficult task, and increased bandwidth requirements by cloud services are a major issue. The problem of combinatorial optimization must be addressed. Task scheduling is required to overcome this issue. Because makespan is influenced by other criteria including the number of tasks, execution time, task scheduling, and fault tolerance, the author gave it priority. ACO is used in this work to reduce makespan by adapting Reinforcement Learning (RL) and fault resistance. The most important tasks are prioritized. These tasks are queued and then mapped to suitable VMs using the Ant Colony Optimization technique. This approach is used to find the optimal solution for optimization problems. Reinforcement Learning is used to avoid local Optima or premature convergence by the ACO algorithm.

**Particle Swarm Optimization (PSO) based Task Allocation**

According to Li et al. [58], there is a load-Balancing issue in different edge nodes that needs to be investigated. The authors presented an edge computing load-balancing strategy that focuses on intermediary nodes. The intermediary node is thought to achieve real-time edge node attributes and classification evaluation. The intrinsic attribute of nodes is stored in intermediary nodes in the architecture, and then load-Balancing techniques are applied. The work takes into account intrinsic attributes such as CPU, physical memory, disc size, network bandwidth, and CPU main frequency multiplied by the number of cores. Another feature is the real-time attribute along with the sample classification set. The experiment results show that the strategy efficiently balances the load between nodes and reduces task completion time when tasks are small; however, the outcome is less efficient when tasks are complex.

**Cost-Aware Big-Bang- Big-Crunch Based Task Allocation**

Big-Bang Big-Crunch technique incorporates bio-inspired analysis approach elements. It follows the characteristics of astrology science. The technique has a high rate of convergence and a low computational cost. This optimization strategy consists of two major steps. Big Bang generates a discrete search space, and Big Crunch congregates the solution at the global optimum value using an expense-oriented fitness value. It has learning abilities derived from bio-inspired evolution. The probabilistic distribution is used to generate populations. Using the Big-Bang phase, the entire population is present in a search space. It gets smaller as it gets closer to the mean point calculated in the Big-Crunch stage, which corresponds to the best global solution [59]. Rawat et al. [60] identified operational cost and efficiency as the primary parameters for measuring cloud QoS. They advocated for efficient resource allocation via task scheduling. For resource allocation, the Big-Bang Big-Crunch-Cost model is provided. Some astrology science features are expected to be followed in BB-BC. The computational cost is low, while the convergence speed is fast. There are two critical steps: Step 1: BB generates a finite search space; Step 2: BC uses a cost-oriented fitness function to converge on the global optimal point. A new population or schedule is produced via optimization at each level, and it depends on the data produced at earlier phases. The population is generated using a probabilistic distribution function. The BB phase is used to distribute the population in search space. During the BC phase, the distributed population size got smaller around the mean point. Task allocation techniques are used to maximize resource utilization. It supports cloudlet mapping from many to one while taking VMs into account. The iteration numbers in this work are not the same. The work takes into consideration optimization techniques, and performance is measured using makespan and resource cost. The optimal solution is provided by the IaaS model, which supports both dynamic and individual task allocation. The results show an improvement in request completion time and average resource enhancement cost, but workflow scheduling in terms of time must be taken into account.

**Shapley Value Based Differential Evolution Algorithm (SVBDEA)**

The Shapley value is a game theory solution concept that calls for equally allocating all gains and losses among several cooperating agents. Game theory describes a strategy that involves two or more players or variables to achieve a desired outcome or payoff. The Shapley value is especially relevant when everyone's efforts are not equal yet everyone tends to cooperate to obtain the benefit or payoff [61]. Differential Evolution (DE) is a straightforward and efficient developmental algorithm for solving global optimization problems in a continuous domain [62]. Ma et al [63] stated that monitoring the quality of service (QoS) provided to users is one of the challenges in the cloud data center as the cloud computing system consists of large-scale servers that are shared with a much larger number of users. Cloud service users have different QoS requirements, and it might be difficult for the service provider to accommodate their demands. To overcome this issue the author developed an algorithm that combines Differential Evolution with the Shapley Value economic model. The algorithm aids in analyzing each VM's function so that tasks can be effectively assigned to resources. This approach uses the "Shapley Value Based DE algorithm (SVBDA)" model which constantly adjusts to customer requirements and solves the issues of different QoS requirements of cloud users. The model showed a significant improvement when compared with the "DE (Differential Evolution)" algorithm and the traditional task-VM binding policy, but other factors like throughput and makespan must be taken into account to assess the task completion time and data center performance.

**Genetic Algorithm based Task Allocation**

Rekha and Dakshayini [64] studied task allocation problems among multi-computing machines and introduced a Genetic based Algorithm for task allocation. The approach aims to reduce task completion time by efficiently allocating resources. Resource Allocation is done by selecting the VM that is suitable, free, and appropriate. Next, the fitness chromosome value is calculated for the given population, to create a new population by repeating the subsequent steps. After completing the process of new population generation, an exchange process is carried out where the newly generated population acts as an existing generation, and the last fitness value is selected for scheduling. The performance of the work is analyzed by comparing the results with Greedy and simple allocation methods based on makespan and throughput. The results indicate an improvement in the system, but energy consumption and resource requirements could also have to be considered for better results.

**Hybrid Algorithm**

Hybrid algorithms combine heuristic and meta-heuristic techniques. Initial VM placement is done using heuristic methods, and VM placement during migration is optimized using meta-heuristic approaches. The meta-heuristic methodology may be used to generate a set of solutions, similar to how a heuristic method is used to discover the optimal solution. When time, money, and available choices are taken into account, both strategies seem attractive; nevertheless, once implementation begins, complexity increases [65].

Hybrid scheduling techniques integrate two planning processes to handle the problem of work allocation for cloud computing [66]. Following are some of the algorithms that have addressed load-Balancing, multi-object job scheduling, and polynomial hybrid frameworks with time class issues.

**Hybrid Genetic Algorithm–Ant Colony Optimization (HGA–ACO) based Task Allocation**

Kumar and Venkatesan [67] stated that task allocation is a polynomial time class issue that is strenuous to get the best solution. This research is an efficient multi-objective HGA-ACO considered for task allocation technique to manage the enormous cloud users' requests. The authors used the utility-based schedule that identifies the task order and the resources that need to be scheduled. HGA-ACO takes the scheduler's output into account and looks for the optimum allocation method, taking response time, throughput, and completion time into account. For the initialization of an efficient pheromone for ACO, the program utilized genetic algorithms. To perform crossover activities, ACO is employed to improve GA solutions.

**Hierarchy Process Algorithm**

The hierarchy process is used for prioritizing incoming tasks. In this process, if the VMs are occupied, the tasks wait until the previously assigned task is completed. The duration and runtime of the job might be used to regulate task priority. The hybrid method executes task scheduling based on parameters including CPU, memory, and storage after the jobs are queued and given priority [68]. Sreenivasulu and Paramasivam [69] identified task scheduling and allocating a suitable resource in the cloud to be a challenging issue. Task scheduling is important so that the VMs can be assigned properly to the task. The author developed a hybrid task-prioritizing approach to address this problem. The framework combines the modified BAT scheduling model with the Bar system concept to prioritize tasks using a hierarchical method. A minimal overload and lease policy is employed for preventive action within the data center to reduce VM overloading.

**Gray Wolf Optimization-Reinforcement Learning (GWO-RIL) based Task Allocation**

Gray Wolf Optimization (GWO) is used to improve task allocation. The algorithm is used to perform accurate workload prediction by using continuous values. GWO is optimized using the Reinforcement Learning (RIL) approach for improving task allocation [70]. Yuvaraj et al. [71] introduced an ML model for job allocation and a correspondent serverless model. The technique is aimed at load-Balancing and approximation output algorithms. In the suggested work, M servers are considered to contain D number of services where cloud resource distribution and energy management processes are considered. A system connected to the cloud is expected to consume energy be it in an active or idle state. M represents the physical array, and D is considered a resource collection in the proposed work. The server assigns VMs based on First come First serve, and accordingly, the resources are allocated. If the resources are occupied, the last needs to wait for the next available resource. Whereas on the local level, the work distribution is continued on and off. The results indicated that GWO-RIL reduces runtime and adjusts to changing load conditions, but the distributed deep learning approach can be employed for task allocation. Moreover, task allocation strategies can be explored for efficient scheduling and task offloading.

**Reference Vector Guided Evolutionary Algorithm-Normal Distribution Angle-Penalty Distance (RVEA-NDAPD)**

Xu et al. [72] stated that the stochasticity, running style, and unexpectedness of user requests in the cloud environment present significant challenges to task scheduling. A task scheduling model is presented as a solution to this problem. The work takes into consideration task characteristics, the user, and the system. RVEA-NDAPD is used in this model for multi-object job scheduling. RVEA divides the objective space into a multitude of subspaces, and classification is performed separately within every subspace. It has been shown that the objective space partition can assist in balancing convergence and variability in decomposition-based techniques. The objective space partition is like adding a limit to the subproblem specified by each reference vector. The generation of tasks is unpredictable and random. Once the tasks are generated, they are combined in a queue and sent for scheduling. The scheduling agents allocate VMs to these tasks, and VM resources are allocated by the resolve center. The allocation strategies appropriate for the framework are obtained as per optimization goals. The number of tasks on VMs is different. Certain task parameters, such as the size and duration of execution, are taken into account when tasks are assigned.

The results show that performance increases, and while many other variables influence the allocation process, task completion effectiveness is the most important one. The effort should concentrate on scheduling model security problems, VM failures, and fault tolerance scheduling.

**Genetic Algorithm and PSO-based Task Allocation**

PSO is an algorithm that looks for the optimal answer among the available options. It is bio-inspired. It simply needs the objective function and is unaffected by the target's slope or any other variable form. Additionally, there are just a few hyperparameters [73]. According to Bharathi et al. [74], load-Balancing is difficult in a dynamic system of workload. To solve the aforementioned issue, the authors devised a two-level scheduling that took into account QoS, resource utilization, and energy efficiency. In this study, PSO and genetic algorithms are coupled to effectively plan jobs on virtual machines. The VMs are mapped to appropriate PMs, which lowers energy consumption. In the initial stage of scheduling, GA and PSO work together to schedule tasks for VMs. It uses a hybrid methodology. The second step focuses on assigning the VM to appropriate PMs by taking the Power Aware Best Fit Algorithm into account. The study showed an improvement in the outcome. However, as scheduling is done through VM mapping and tasks are relocated, it has to take into account factors that can help in preventing security attacks.

**Artificial Bee Particle Swarm Algorithm-based Task Allocation**

The Artificial Bee Colony (ABC) was initially presented by Karaboga in 2005 [75]. This algorithm's main benefit is that exploration and exploitation are quite well balanced. Exploitation employs well-known methods to minimize a cost function, whereas exploration browses the search space for fresh ideas in unexplored areas. According to Maheswari et al. [76], work scheduling in a cloud environment plays a crucial role and therefore influences performance. Task scheduling is an NP-hard issue since there are several possible solutions. The authors want to reduce makespan, and cost, and maximize resource utilization while effectively balancing the load. Particle swarm and artificial bee colony algorithms are combined to accomplish these goals. In PSO, the local search weakness is the main drawback. Al-Maamari et al. [77] suggested that the problem of makespan, high cost, and resource utilization occurs due to the possibility of getting locked in local search when the last iteration takes place. An effective task-scheduling technique was developed to address this problem. In this system, T is a collection of user-submitted tasks, T1 through Tn, and VM is a set of virtual machines, VM1 through VMn. The allocated data center employee bee keeps track of VM maintenance information in the data center's hive table. The population of Scout Bees is taken to be as T, and initializing food source and calculated fitness Value are represented by VM and I. For each VM employee, the bee updates the hive table. The hive table keeps track of information including VM cost, RAM and CPU availability, VM load, processor speed, storage, bandwidth, and processor MIPS. This information is stored in sorted order. Based on the information in the hive table, the partial best fit (p best) for the job at hand is determined. The table's p Best and g Best values are updated if the fitness proves to be superior to the prior best fit and superior to the previous global best value, respectively. The VM is chosen depending on its workload, processing speed, and cost. The findings show that while cost and turnaround time have improved, security issues have not been taken into account in this effort.

**RQ.4. What are the parameters that were unaddressed related to task allocation in the data center?**

The improper task allocation has been one of the reasons for high energy consumption in cloud data centers. The major challenge for [48] was to design an algorithm for binary computation offloading which could reduce energy consumption in cloud data centers. The work showed great service and task allocation efficiency as compared to single resource allocation. Additionally, the results showed that the algorithm accomplishes a trade-off between time complexity and optimality loss. The algorithm achieved a balance of performance and speed for multi-resource allocation, although task allocation in complex systems still requires improvement.

The major challenge for [49] was to design a system that could deal with heavy workloads and keep energy consumption in check. The findings showed that EFRO increased the energy efficiency of the Round Robin (RR) and Most Efficient Server First (MESF) schemes by 54.83% and 71.28%, respectively. Instead of only allocating one resource, the task concentrated on allocating several resources. The algorithm designed by [51] indicated that the energy consumption got improved by approximately 14%, 6.3%, and 2.8% in comparison to random allocation algorithms like Cloud Z-Score Normalization and multi-objective scheduling algorithm with Fuzzy resource utilization.

The SLA violation was observed under different situations. In comparison to the previously mentioned methodologies, it was shown that when VMs differ, the average energy usage improves by 10%, 5%, and 1.2%, respectively. However, there was no change in makespan, and the work needed to focus on priority-oriented tasks.

The challenge for [52] was to design a low-complexity algorithm for resource management. The results indicated that the system performance was improved and could be considered for combining computational offloading with low signaling overhead. The work neglected the parameters like bandwidth and energy cost of offloading.[25] stated that there is a requirement for a system that improves task allocation and resource scheduling. The authors gave BATS and Bar optimization algorithms to solve this issue. The results were compared with the BATS and Improved Differential Evolution Algorithm (IDEA), and it was observed that the system reduces response time to 50%, but the work still needs to focus on effective scheduling. In addition, the framework didn't consider pre-emption, and the tasks considered were of the same size.

The major challenge for [57] was to design a task allocation algorithm that could meet the growing bandwidth requirement. According to the observations, Reinforced-Ant Colony Optimization (RACO) outperforms ACO by 60%. The fault tolerance was enhanced by 10%, thus improving the performance of the system. The work was suitable to find the optimal solution for task scheduling, but the concept of load-Balancing, QoS parameters, and cost minimization could be used in the author's work. The system needs to be tested with real-time data.[58] stated that the strategy works better when the tasks are small. [60] indicated that Big-Bang Big-Crunch Cost improves finishing time by 15.23% when user requests were 300, and the average time was improved by 19.18% when a user request is 400. The resource cost is enhanced by 30.46% when the population is 400. The work could focus on workflow scheduling considering time and resource costs. The SLA inclusion and energy-saving parameters could improve the performance of the work.

The major challenges for [63] were to design a system that meets varying QoS requirements of the users. On comparing the results of work with the Differential Evolution (DE) algorithm and the conventional task-VM binding policy, the Shapley value-based DE algorithm (SVBDA) indicated a huge improvement. The model adapted to user needs in terms of cost, bandwidth, and execution time dynamically.[64] considered the task allocation issues among multi-computing machines and gave a Genetic-based algorithm. The results show that the system had improved, but for even better outcomes, energy usage and resource requirements could be taken into account.

By comparing the response time of GA-ACO with GA and ACO algorithms, [67] found that the algorithm's time was minimized by 140 and 100 ms. When taking into account the completion time, they found that the time is minimized by 144 and 117 ms. On comparing the throughput, they concluded that the method performs 9% and 6% better than the other two algorithms. The major challenge for [69] was to design an algorithm that focused on task scheduling and resource allocation. The results indicated that the response time for the system was reduced to 45%, and the execution time was also reduced as the utilization of memory increased. There was a 30% improvement in memory utilization, but the work needs to be evaluated considering the real-time workload.[71] gave an ML model for job allocation. The results indicate that throughput was 32% efficient when the load was 30. Latency was 850 ms when the load was taken as 30. The cost was less in comparison to the other algorithms. The challenge for [72] was stochasticity, running style, and unexpectedness of user requests. They introduced a task allocation model to resolve this challenge, and the results indicate that the performance got improved, and a suitable task allocation strategy was obtained. But the work needed to focus on security issues in the scheduling model, mainly on VM failures and fault-tolerant scheduling.

By eliminating SLA violations, [74] were able to accomplish effective load-Balancing, resource utilization,

and QoS. However, the security of the cloud environment was not taken into account. [76] considered job scheduling an important part of the cloud environment. To overcome this issue, a job scheduling-based ABPS model was given. When makespan was taken into account, the ABPS model outperformed the Artificial Bee Colony algorithm (ABC) and Particle Swarm Optimization (PSO) by 22.07% and 28.12%, respectively. When compared to ABC and PSO, ABPS demonstrated improvements in resource usage of 49.37% and 48.88%. Although the imbalance was decreased to 16.21% and 20.51%, task security was not taken into account.

**RQ.3. What are the role of performance and its evaluation in cloud data centers?**

## 5. Performance Management

In a cloud system, performance management is the process of analyzing multiple matrices and benchmarks. The effectiveness of each application must be evaluated and compared to that of other applications or techniques. Due to the dynamic nature of the cloud paradigm, resource management in the cloud is difficult. Performance metrics are used to analyze the attributes to identify what factors affect performance [78].

### 5.1. Performance Metrics

Performance management of the cloud framework is assessed using the response time, overall SLA violations, and electricity utilized by the resources to handle user requests. To minimize the detrimental effect on QoS, it is common practice to incorporate specific parameters and performance metrics as the goal function [79,80]. Performance measurements can provide important details for identifying the real cause of a breakdown that spreads across several system layers [81, 82, 83]. Numerous factors, including CPU, memory, and network utilization, may be used to assess the cloud data center's performance. When it comes to task allocation, other factors such as task priority, nodes, dependability, availability, throughput, system response time, workload, energy, execution time, QoS, and SLA are taken into consideration [84,85,86]. Several research, including [87,88,89,90,91], divided the performance evaluation criteria into task- and resource-based categories. The characteristics taken into account for the task are reliability, availability, bandwidth, cost, energy temperature, utility, and workload, while the parameters taken into account for the resources are CPU, network, VM, node, and storage.

### 5.2. Performance measures

It has been observed in recent years that analyzing the performance of cloud environments now requires a thorough examination of data center performance. The performance of the data center is critically dependent on several factors, including the cooling process, information technology services, and management of the power supply, along with productivity and efficiency [92]. Some of these performance parameters are discussed below:

**Computer Processing Speed:** The computational power is expressed in cycles per kilowatt-hour. By contrasting it with the energy required to operate the system, the power efficiency of the data center can be assessed. Therefore, the data center administrators assess the exact amount of energy consumed to verify the accurate energy intake [93].

$$CP_{DC} = \int_{t_1}^{t_2} E(t)\, dt \tag{1}$$

**Power Usage Effectiveness (PUE):** It is the ratio of total power consumption and power utilized by IT equipment. PUE has emerged as the industry's ideal metric for evaluating the data center frameworks for energy efficiency [94]. In Equation 2, PUE is the Power Usage Efficiency which is calculated by the total power consumed by data center TDC over the total power consumed by the equipment TIT.

$$PUE = \frac{\sum T_{DC}}{\sum T_{IT}} \tag{2}$$

**Carbon Emission (CE):** Carbon emission measures the quantity of CO2 released within the data center. The total amount of energy consumed is converted to CO2 emissions using the Carbon Emissions Factor (CEF). The inclusion of different power sources, such as carbon, gas, wind, solar, biomass, nuclear, etc., which also have an impact on overall electrical output and conversion efficiency, determines CEF [95].

This factor, therefore, varies from one data center to the next. In Equation 3, CE is the carbon emission which is calculated by total energy (TDC) consumed by the data center into Carbon Emissions Factor (CEF).

$$CE = TDC * CEF \qquad (3)$$

**Productivity:** The data center's productivity depends on how it manages the necessary services. Productivity is also influenced by who is running these services. Owners are required to use the data center resources more effectively and to adhere to SLA if they sublet their services to resellers. According to Equation 4, productivity in a cloud data center may be defined as the user resource request time over the request execution time [96-100].

$$Productivity = User\ Request\ Resource\ Time\ /\ Request\ Execution\ Time \qquad (4)$$

### 5.3. Performance Management Approaches

In this work, we examine various approaches proposed by researchers to address the challenges of dynamic allocation in cloud environments. Addis et al. [101] present a resource allocation strategy aimed at handling multi-tier cloud systems and maximizing revenue while ensuring multi-class SLA satisfaction. The authors utilize power and control theory models to tackle the problem of resource allocation at very small-time scales. They propose a local search-based heuristic approach that guarantees the availability of running applications. The primary objective of their initiative is to investigate resource allocation strategies for virtual cloud environments that can evaluate performance and energy trade-offs. Although their approach efficiently manages all requests, it has limitations in terms of scalability and the consideration of energy consumption.

Another approach, discussed by Guazzone et al. [102], focuses on power-aware resource management to achieve suitable Quality of Service (QoS) and reduce energy usage in cloud systems. Their framework aims to optimize the allocation of physical machines (PM) and virtual machines (VM) to minimize SLA violations and maximize profits. By analyzing and retaining application performance goals, the authors propose an automatic framework for computing resources. The performance metrics considered in their approach include power consumption and efficiency. The results indicate that their approach dynamically adapts to changing workloads, reducing QoS violations and energy consumption. However, their work lacks a focus on migration management and resource negotiation between physical sources and the cloud environment.

Yuan and Liu [103] address the challenge of low utilization of pre-reserved resources and cost management in cloud environments. They investigate a dynamic performance management strategy that incorporates resource borrowing and lending strategies. By considering pre-reserved resources and employing resource management strategies, their approach aims to optimize resource utilization. The performance metrics for their approach include resource pre-reservation consumption time. The results demonstrate that their model outperforms in scenarios where resource lending or borrowing is required. However, one limitation of their approach is the absence of a system to determine resource updates.

Fareghzadeh et al. [104] propose a cutting-edge technique called Dynamic Performance Isolation Management (DPIM) for performance isolation control in cloud environments. Their method utilizes an architectural framework that enables service providers to deploy multiple isolation techniques and enforce performance isolation silently. The authors emphasize the usefulness of this performance isolation management technique and its related framework in various service scenarios. The performance metrics considered in their approach include CPU utilization and throughput. The results indicate the integration of several aspects and a performance improvement. However, their model is not evaluated in diverse scenarios, such as real-time environments.

Younge et al. [105] introduced a power-based scheduling framework for green computing. Their approach aims to manage resources efficiently, minimize virtual machine (VM) designs, enable live migration, and enhance system efficiency. The authors address the problem of energy consumption and $CO_2$ emissions by data centers. Their power-based scheduling framework considers power consumption as a performance metric. The results indicate that the system achieves energy savings. However, one limitation of their work is the absence of power and thermal awareness to further increase energy savings for physical servers.

Yamini [106] highlights task consolidation as a significant approach for streamlining resource utilization and improving energy efficiency in cloud environments. The author develops two energy-conscious task consolidation heuristics after establishing a link between resource use and energy consumption. The problem addressed in this work is global warming and carbon emissions associated with cloud computing. The proposed heuristics are evaluated based on a green algorithm that focuses on energy consumption. The results indicate improvements in energy savings and potential savings in other operational costs. However, this work only considers the energy aspect and other parameters such as execution and workload could be further explored.

Chen and Li [107] introduce a queueing-based methodology for cloud performance management, specifically targeting web applications. They utilize queueing theory to dynamically generate and delete virtual machines (VMs) as service centers to enable scaling up and down. The proposed model eliminates the need for live VM migration, simplifying the process. The problem addressed in this work is resource scaling. The performance metric used is response time, and the results indicate the effectiveness of the model in scaling up and down. However, further experiments are needed to precisely measure the impact of the model on the usage of computing resources.

Sun et al. [108] present an architectural model for resource management and monitoring by combining the power of Service-Oriented Architecture (SOA) and IT service management concepts. The system utilizes an Alarm Rule Configuration algorithm and a Web Service basic Model. It accesses and monitors data for conducting Quality of Service (QoS) analysis. The proposed hierarchical model ensures isolation between users and physical resources. However, the system's cost and complexity are higher compared to other approaches.

Puviani and Frei [109] describe two adaptation patterns and propose a switching mechanism based on system conditions. Under normal conditions, a peer-to-peer (P2P) structure with inherent adaptability is adopted. However, in stressed situations with failed nodes and increased service demands, the system transitions to a temporary centralized structure. The live nodes negotiate to establish temporary management. The problem addressed here is user dynamic requests in cloud-based applications. The performance metric considered is self-* (self-adaptation). The simulations demonstrate the promising nature of the P2P approach, but further research is needed for real-world implementation.

Nahir et al. [110] investigate the effectiveness of service providers who distribute work among private and cloud resources. They suggest assignment techniques for large-scale distributed systems to enhance system performance. The resource allocation technique balances the end-user experience and operational costs of renting resources from the cloud provider. The problem addressed is resource management and allocation. The proposed approach is based on game theory, and the performance metric considered is load changes. The achievement of sharply changing load is highlighted. However, this work focuses primarily on scaling and further exploration of other aspects is warranted. Maurer et al. [111] propose a novel technique called MAPE (Monitoring, Analysis, Planning, Execution) for monitoring the cloud and facilitating SLA (Service Level Agreement) management. The technique incorporates low-level system matrices and SLA parameters to balance the virtualization layer and improve the monitoring process. The planning and analysis phases are integrated with knowledge-based management (KM). The KM phase monitors system information and takes reactive actions to prevent SLA violations. The problem addressed in this work is cloud infrastructure management. The performance metrics used are utilization and RAE (Resource Allocation Efficiency). The achievement of the work is efficient management of infrastructure, energy, and scalability. However, other parameters such as energy consumption and task execution time have not been considered.

Grolinger et al. [112] focus on the storage component of cloud computing and address challenges related to data management. They propose solutions that consider data structures, querying, scalability, and security. The problem addressed is the diversity and inconsistency of terminology, limited documentation, sparse comparison and benchmarking criteria, occasional immaturity of solutions, lack of support, and the absence of a standard query language. The performance metrics used are scalability, fault tolerance, and availability. The achievement is to identify and select proper storage solutions. However, inconsistent support, limited documentation, and lack of standardization remain limitations.

Fargo et al. [113] identify three major research categories for task allocation and seven major performance management approaches in cloud computing. They highlight research gaps within each approach and suggest new perspectives to bridge these gaps. The problem addressed is power consumption in data centers and cloud systems. The proposed approach is an autonomic power and performance management method called Autonomic Workload and Resource Management (AWRM). The performance metrics used are VM power and power reduction. The achievement is an improved management system.

Guan et al. [114] develop an unsupervised failure recognition method that utilizes Bayesian models to identify system anomalies. System administrators confirm and categorize the anomalies, and supervised learning based on decision tree categorization is used to prevent future failures. The problem addressed is a system failure. The proposed approach utilizes machine learning techniques to solve this issue. The performance metric used is the true positive rate. The achievement is a high true positive rate and a low false positive rate for failure recognition. However, other failure management techniques could be considered for prediction.

Gupta et al. [115] focus on Virtual Network Services (VNS) and the challenges of providing services with performance and availability similar to conventional networks. They incorporate machine learning techniques, including deep learning and ML, to detect and localize faults and improve the performance of VNSs. The problem addressed is latent faults and performance in a multi-cloud environment. The performance metrics used are time and correctly classified instances. The achievement is the effective handling of fault issues. However, real-time experimentation is required to validate the approach.

Jhawar et al. [116] propose a framework that considers security, reliability, and availability in the context of resource allocation in cloud infrastructure. They go beyond traditional performance/cost-oriented resource consumption and take into account security criteria defined by users and providers. The problem addressed is security, reliability, and availability in the cloud environment. The proposed approach is a resource capacity-based approach. The performance metrics used are the reserve list and allocation. The achievement is a secure system. However, only latency is considered, and other aspects of security, reliability, and availability could be explored.

Ghamkhari et al. [117] present a systematic strategy to increase customer profit from green data centers by considering local renewable energy generation and the stochastic nature of workloads. They propose an optimization-based data center profit maximization technique for scenarios with and without renewable generators. The problem addressed is the trade-off between reducing energy use in data centers and increasing revenue from the internet and cloud computing services. The performance metrics used are server utilization, profit gain, and loss probability. The achievement is an improved cost factor. However, the model is based on assumptions, and real-world validation is needed. Gulati et al. [118] suggest the Pesto algorithm for congestion and load-Balancing in heterogeneous virtualized data centers. The algorithm automates storage performance management for load-Balancing and VM decentralization. The system constructs black box performance models for storage devices. The problem addressed is the challenging estimation of IO performance on shared storage. The performance metrics used are latency and throughput. The achievement is load-Balancing and effective models. However, the lack of storage remains a limitation.

Guazzone et al. [119] introduce a framework for resource management to achieve suitable Quality of Service (QoS) and reduce energy consumption. The framework dynamically adapts to changing workloads, reducing QoS violations and energy consumption. The problem addressed is the reduction of QoS violations and energy utilization. The proposed approach is a framework that automatically manages resources. The performance metric used is the best fit for decreasing energy consumption. The achievement is significant improvements in performance and energy consumption. However, real-time experimentation is required for validation.

Singh et al. [120] discuss the use of model predictive control (MPC) for cloud resource allocation to address the performance degradation caused by increasing web service traffic. The MPC controller optimizes the allocation of resources, such as VMs, to meet response time and resource quantity constraints. The problem addressed is the degradation of system performance due to increasing web service traffic. The performance metrics used are response time and CPU utilization. The achievement is better resource utilization. However, if the MPC fails, performance degradation may occur.

### 5.4 Comparative Study of Performance Management Approaches: Findings and Discussions

As the requirement for cloud computing is increasing, new opportunities for improved performance awareness systems are also increasing. Performance management gives the idea about the health state and performance of the system [121]. The approaches discussed in section 5.2 give various techniques that help in performance management in the cloud environment. As [101] stated, changing demand of the cloud users is a challenge for service providers as QoS can't be achieved, thus affecting the overall performance of the cloud environment. The author introduced a resource allocation policy to meet the changing demands of service users. For the simulation, the comparison service centers that had the capacity of 400 servers and 50 requests were managed efficiently. Resource allocation issues were handled by adopting a control theory model, but the system is not scalable, and energy utilization of servers was not considered. The major challenge for [102] was QoS violation issues. To resolve this issue, the author introduced a framework for managing computing resources. The system was compared with approaches STATIC_SLO and STATIC Energy. The results indicated that energy utilization is lower than both approaches. SLA violation is under 1%. It can be concluded that VMs can be consolidated on the same PMs, hence reducing energy consumption.

A dynamic performance management strategy was investigated by [103]. The author introduced a resource management strategy. The experiment results indicated performance improvement. Resource Per Reservation (RPR) focuses on resource pre-reservation strategies, and Resource Borrowing/Lending (RBL) is considered a lending/borrowing strategy, but the work introduced by the author shows improvement in request handling and resource utilization. [104] considers various execution techniques for different stages of operating entities in varying service interfaces, which lacked in other related studies. Performance isolation is discussed in terms of SLA, and multi-level isolation is discussed as a fundamental cloud service. The major challenge for [105] was to design a framework that reduces the energy consumption and $CO_2$ emission by the data center. The author introduced a power-based scheduling framework to solve the issue. The results indicated that there was a 12% conservation of power consumption. [106] also considered the same challenges as [105] for their framework. The results indicated that there is 13-18% energy saving within the system when migration was considered. It was concluded that the task with low source utilization is most preferred for consolidation.

The challenge for [107] was resource scaling, and to overcome this issue, a queueing-based model was introduced. The result showed that the model is effective for scaling up and down. However, the experiments are not enough to precisely measure the effect of the model on the usage of computing resources. [108] introduced a model to achieve isolation between the machine and the user, which helps in changing operations from passive to active mode. In actual conditions, the model is complex. A simulation was conducted by [109], and the results indicate that P2P is a promising approach as each mode performs different roles, i.e., independent peer, temporary manager, and temporary follower. When the system goes under stress, the nodes fail. The challenge for [110] was resource allocation and management to design a distributed approach. The work depended on actual task execution time, hence reducing the queuing overhead in the server. The results indicate that the replication scheme improves the selection of the processing server, and the queuing overhead is improved. The major issue for [111] to design the novel technique for cloud monitoring was cloud infrastructure management. During the study, the Case-Based Reasoning (CBR) indicated that the SLA violation was reduced to a third as compared to no CB. The resource utilization was high, and CBR response to user requests was 12 seconds, whereas VM took 0.24 seconds to perform decision-making.

The storage and data management issues on the cloud were studied by [112]. The solution given was considered an alternative to traditional relational databases that handle a large amount of data. The major challenge for [113] to design a power-optimized system was to understand the cloud's power requirements and accordingly manage them. The comparisons between the approach and the static resource allocation strategy, the adaptive frequency scaling strategy, and a similar multi-resource management strategy showed that the approach may reduce power usage by up to 87%, 72%, and 66%, respectively. System failure has been a concerning issue for [114]. The author gave an ML approach to solve this issue. The experiment conducted indicated that the system obtained a high true positive rate and a low false positive rate for proactive failure management. [115] suggested that Support Vector Machine's (SVM) performance was better than the Random Forest algorithm with >=95% accuracy. The true positive rate was high, and the failure rate was low. [116] introduced an approach for identifying various requirements and introduces a heuristics-based approach that considers VM allocation to external hosts. The major challenge for [117] was energy consumption in the data center. The author suggested a profit maximization approach. The experimental results show that, although the model was assumptions-based, the performance of the approach optimization-based profit maximization technique greatly surpasses two comparable energy and performance management algorithms. [118] conducted an experiment using the Pesto algorithm to demonstrate that the system adapts to changes quickly, and the workload performance gets improved by 19%. The experiment also indicates that the throughput improves by 10%, and load latency was minimized by 19%. A challenge for [119] was QoS violation and minimizing energy utilization. The author introduced a framework to overcome this issue, and the results indicated that VM migration improved performance and energy consumption, but the work lacks the real-time experimentation required. [120] used a realistic Web service testbed finding revealed that the controller could meet the required response time restriction even when the service was subject to workload surges and interference, but performance degraded if MPC failed.

## 6. Results

The systematic literature survey results are organized based on the research questions listed in Tab. 1. It has been established that, in the current literature review, 35.72% of the articles focusing on task allocation are published in prestigious journals as well as at distinguished conferences and workshops (shown in Fig 2). Besides that, 26.06% of the papers on performance management were found. Furthermore, several papers on cloud data centers, task allocation, task allocation techniques, performance management, and performance management techniques have been published in the Journal of Supercomputing and IEEE Transactions on Service Computing. We also discovered that Springer publishes 6.98% of research items, ACM publishes 12.41%, Elsevier publishes 14.34%, and IEEE contributes 17.44% of the total research done in this area.

**Table 4**. Articles that achieved Research Questions

| Author | RQ 1 | RQ 2 | RQ 3 | RQ 4 |
|---|---|---|---|---|
| Jim [2] | √ |  | √ |  |
| Xu and Palanisamy [3] | √ |  |  | √ |
| Akhter and Othman [4] | √ |  | √ | √ |
| Usman et.al [5] | √ |  | √ | √ |
| Kumar and Parthiban [6] | √ | √ | √ | √ |
| Shelar et.al [7] | √ | √ |  | √ |
| Rathor et.al [8] |  | √ | √ |  |
| Khurana and Marwah[9] |  | √ | √ | √ |
| Banga et al. [10] | √ | √ | √ | √ |
| Rodriguez and Buyya[11] |  | √ |  | √ |

| Author | RQ 1 | RQ 2 | RQ 3 | RQ 4 |
|---|---|---|---|---|
| Weerasiri et al. [12] | √ | √ | √ | √ |
| Khallouli and Huang [13] | √ | | | |
| Singh and Chana [14] | √ | | √ | √ |
| Singh and Chana [15] | √ | | √ | √ |
| Jennings et al. [16] | √ | | √ | √ |
| Jiang [17] | √ | √ | √ | √ |
| Mann et al [18] | √ | | √ | |
| Milani and Navimipour[19] | √ | | | √ |
| Ramezani et.al. [20] | √ | | √ | √ |
| Zhou et.al.[21] | √ | | √ | √ |
| Rahman et al [22] | √ | √ | √ | √ |
| Krishna [23] | √ | √ | | √ |
| Akintoye and Bagula [24] | | √ | √ | |
| Gawali and Shinde [25] | | √ | √ | √ |
| Hussain and Bagh [26] | √ | √ | √ | √ |
| Mor et al [27] | | √ | | √ |
| Jain and Upadhyay [28] | √ | √ | √ | √ |
| Singh and Kumar [29] | √ | | | |
| Rodrigues et al [30] | √ | | √ | √ |
| Dubey and Sharma [31] | √ | | √ | √ |
| Mohanty et.al [32] | √ | | √ | √ |
| Mehrotra and Srivastava [33] | √ | √ | √ | √ |
| Chen and Li [34] | √ | | √ | |
| Zareian et al [35] | √ | | | √ |
| Gupta et al [36] | √ | | √ | √ |
| Song [37] | √ | | √ | √ |
| Ganesh et.al [38] | √ | √ | √ | √ |
| Fareghzadeh et. al [39] | √ | √ | | √ |
| Van et.al [40] | | √ | √ | |
| Akhter and Othman [42] | | √ | √ | √ |
| Bilal et al [43] | √ | √ | √ | √ |

| Author | RQ 1 | RQ 2 | RQ 3 | RQ 4 |
|---|---|---|---|---|
| Rahman et al[44] |  | √ |  | √ |
| Xu et al[45] | √ | √ | √ | √ |
| Rashid and Parvez [46] | √ | √ |  |  |
| Almezeini and Hafez[47] | √ | √ | √ | √ |
| Liu et. al [48] | √ | √ | √ | √ |
| Gai et. al [49] | √ | √ | √ | √ |
| Wang and Yu[50] | √ | √ | √ | √ |
| Mishra et. al [51] | √ | √ | √ |  |
| Josilo and Dan [52] | √ | √ |  | √ |
| Lin et.al [53] | √ |  | √ | √ |
| Acedo and Rosa [54] | √ |  | √ | √ |
| Liao et.al [55] | √ | √ | √ | √ |
| Kaelbling et. al [56] | √ | √ |  | √ |
| Nalini and Khilar [57] |  | √ | √ |  |
| Li et. al [58] |  | √ | √ | √ |
| Rawat et. al [60] | √ | √ | √ | √ |
| Ma et. al [63] |  | √ |  | √ |
| Rekha and Dakshayini [64] | √ | √ | √ | √ |
| Kumar and Venkatesan [67] | √ | √ |  |  |
| Makwe and Kanungo [68] | √ | √ | √ | √ |
| Sreenivasulu and Paramasivam [69] | √ | √ | √ | √ |
| Yuvaraj et. al [71] | √ | √ | √ | √ |
| Xu et. al [72] | √ | √ | √ | √ |
| Maheswari et al. [76] | √ | √ | √ | √ |
| Al-Maamari et al [77] |  | √ |  | √ |
| Addis et.al [92] | √ | √ | √ | √ |
| Guazzone et.al [93] | √ | √ |  |  |
| Yuan and Liu [94] | √ | √ | √ | √ |
| Fareghzadeh et.al [95] | √ | √ | √ | √ |
| Younge et. al [96] | √ | √ | √ | √ |
| Yamini [97] | √ | √ |  |  |

| Author | RQ 1 | RQ 2 | RQ 3 | RQ 4 |
|---|---|---|---|---|
| Chen and Li [98] | √ | √ | √ | √ |
| Sun et .al[99] | √ | √ | √ | √ |
| Puviani and Frei [100] | √ |  | √ | √ |
| Nahir et.al [101] | √ | √ | √ | √ |
| Maurer et. al [102] | √ | √ | √ | √ |
| Grolinger et. al [103] |  | √ |  | √ |
| Fargo et.al [104] | √ | √ | √ | √ |
| Guan et. al [105] | √ |  |  |  |
| Gupta et.al [106] | √ |  | √ | √ |
| Jhawar et.al [107] | √ | √ | √ | √ |
| Ghamkhari et. al [108] | √ | √ | √ | √ |
| Gulati et.al [109] | √ | √ |  | √ |
| Guazzone et. al [110] |  | √ | √ |  |
| Singh et. al [111] |  | √ | √ | √ |

## 7.Open Issue

As power costs and carbon emissions have increased, power usage reduction has become a problematic issue for cloud computing [4,17]. Task distribution that considers energy use is essential for reducing data center energy use. There is a demand for methods that take less time to compute and execute. There are possibilities for efficient allocation practices in this research area in the context of appropriate placement opportunities. This is very important with greater influence and implication-based opportunities. Energy consumption should not only be considered for the servers but for computing machines as well. The use of a Hybrid data center architecture using optimal techniques can be explored for energy efficiency, robustness, and scalability issues. The future work could also be in the direction of the parameter's variation in the energy-efficient algorithm, simulation parameters of the cloud, and task requirement-based selection.

The use of heterogeneous evaluation criteria in the reviewed literature poses a limitation. Different studies may have employed different metrics to evaluate task distribution and performance management techniques. Future research should strive to adopt standardized performance measures to address this issue. Standardization would improve the comparability of techniques and enable better benchmarking, leading to more reliable and meaningful results.

**Author Contributions:**

Nidhika Chauhan: conceived and designed the study, collected and analyzed the data, and led the entire research process.

Navneet Kaur: provided valuable inputs during the manuscript preparation.

Kamaljit Singh Saini: support in the research process.

Sahil Verma: provided guidance in various aspects of the study and research process.

Abdulatif Alabdulatif: provided insights and support during the research process.

Ruba Abu Khurma: contributed to the research unit and offered valuable insights into the study.

Maribel Garcia: : provided insights and support during the research process.

Pedro A. Castillo: provided guidance and mentorship throughout the study, and reviewed the results.

**Acknowledgement:** I thank Chandigarh University for providing me with the opportunity to study as a research scholar. We thank the editor and reviewers for their guidance and valuable suggestions. We are thankful to our colleagues and family for always being there to guide and support us.

**Funding Statement:** We are also grateful to Woosong University Academic Research Fund, Korea for their support and funding for our work. This work was supported by the Ministerio Español de Ciencia e Innovación under project number PID2020-115570GB-C22, MCIN/AEI/10.13039/501100011033 and by the Cátedra de Empresa Tecnología para las Personas (UGR-Fujitsu).

**Conflicts of Interest:** The authors declare that they have no conflicts of interest to report regarding the present study.

**Availability of Data and Materials:** The data that support the findings of this study are available from the corresponding author, PAC, upon reasonable request.

**Appendix A. A Criteria for Quality Assessment**

| Criteria | Yes | No |
|---|---|---|
| **Part 1: Question for Screening papers** | | |
| Is the paper related to task allocation in cloud data centers? | | |
| Is the paper related to the performance of cloud data centers? | | |
| **Part 2: Question for Screening papers** | | |
| Is the study aiming to focus on task and performance-related issues? | | |
| Did any subcategories focus on the paper? | | |
| **Part 3: Detailed Questions** | | |
| Was the data mentioned in the paper apt for comparison? | | |
| Are the important parameters for comparative analysis specified? | | |
| Is the study considered explicitly? | | |
| Did the study mention how the system and subject were identified and selected? | | |
| **Part 4: Detailed Questions** | | |
| Did the study mention types of task allocation? | | |
| Did the study mention types of performance management approaches? | | |
| How efficiently is task allocation classified? | | |
| Did the study mention the type of tool used or is it inferred from the study? | | |
| Tools Used | | |
| Was the tool used specified? | | |
| Did the author develop a new tool or utilize existing tools for analysis? | | |